\documentclass[
reprint,
superscriptaddress,
amsmath,amssymb,
aps,
prb,
floatfix,
]{revtex4-2}

\usepackage{graphicx}
\usepackage{mathtools}
\usepackage{dcolumn}
\usepackage{bm}
\usepackage[dvipsnames]{xcolor}
\usepackage[colorlinks=true,linkcolor=blue]{hyperref}
\usepackage{graphicx}
\usepackage{multirow}
\usepackage{lipsum,booktabs}
\usepackage{amsmath}
\usepackage{enumerate}

\begin{document}

\title{Effects of antiferromagnetic coupling and pinning on domain wall dynamics in synthetic ferrimagnets}

\author{Sougata~Mallick}%
\email{sougata.physics@gmail.com}
\affiliation{Laboratoire Albert Fert, CNRS, Thales, Université Paris-Saclay, 91767 Palaiseau, France}
\affiliation{Department of Physics and Nanotechnology, SRM Institute of Science and Technology, Kattankulathur- 603203, Tamilnadu, India}%

\author{Nicolas~Reyren}%
\affiliation{Laboratoire Albert Fert, CNRS, Thales, Université Paris-Saclay, 91767 Palaiseau, France}%

\author{André~Thiaville}%
\affiliation{Laboratoire de Physique des Solides, CNRS, Université Paris-Saclay, 91400 Orsay, France}%

\author{Philippe~Ohresser}%
\affiliation{Synchrotron SOLEIL, L'Orme des Merisiers, Saint-Aubin - BP 48, 91192 Gif-sur-Yvette, France}%

\author{Nicolas~Jaouen}%
\affiliation{Synchrotron SOLEIL, L'Orme des Merisiers, Saint-Aubin - BP 48, 91192 Gif-sur-Yvette, France}%

\author{Vincent~Cros}%
\email{vincent.cros@cnrs-thales.fr}
\affiliation{Laboratoire Albert Fert, CNRS, Thales, Université Paris-Saclay, 91767 Palaiseau, France}%

\author{Vincent~Jeudy}%
\email{vincent.jeudy@universite-paris-saclay.fr}
\affiliation{Laboratoire de Physique des Solides, CNRS, Université Paris-Saclay, 91400 Orsay, France}%

\date{\today}

\begin{abstract}
Domain wall (DW) dynamics in antiferromagnetic (AFM) systems offer the advantages over their ferromagnetic counterparts of having faster and more energy efficient manipulation due to the absence of net magnetization, leading to reduced magnetic crosstalk and improved performance in spintronic devices. A comprehensive analysis of DW dynamics across regimes such as creep, depinning, and flow is well established in ferromagnetic systems but remains lacking in AFM-coupled systems. In this study, we explore the nature of DW dynamics in synthetic ferrimagnetic multilayers composed of Pt$|$Co$|$Tb$|$Al for different Tb thickness, focusing on the underlying pinning parameters, and on the different regimes of DW dynamics driven by spin-orbit torques (SOTs). We find that due to the AFM coupling between Co and Tb, the magnetic moment of Tb increases with Tb thickness resulting in a reduced saturation magnetization and an enhanced depinning field. The DW disorder interaction is found to vary weakly with the AFM coupling between Co and Tb, while the complete withdrawal of the Tb layer strongly increases the anisotropy and the DW pinning. Furthermore, we propose a novel approach to measure effective SOTs by comparing depinning transitions in current- and field-induced DW motion. This research reveals new insights into DW dynamics in coupled AFM systems, highlighting enhancements in mobility through optimized SOTs and pinning landscapes.
\end{abstract}

\maketitle

\section{\label{sec:intro}INTRODUCTION}

Controlling the motion of magnetic domain walls (DWs) is crucial for understanding and manipulating the magnetic properties and spintronic behavior of devices in a wide range of applications, such as data storage, logic elements, sensing, and neuromorphic computing \cite{Kumar2022_DWMemory,Allwood2005_DWLogic,Parkin2008_DWRacetrack, Aroop_racetrack_JPD}. The performance and reliability of DW-based devices are inherently tied to the speed of DW motion, as enhanced mobility reduces the energy cost and facilitates the read and write operation \cite{Kim2017_FastDWFerri,Miron2011_FastDWRashba, DW_Relax_Mallick_Sci.Rep.}. A first way to move DWs is through torques generated by external magnetic field pulses. In this case, the efficiency of DW motion, notably in the lower range of field amplitude, primarily depends on the pinning landscape as it might induce DW roughness, deformation and stochasticity of motion \cite{Gorchon2014_PinningDW,Jeudy2016_UniversalPinning,Jeudy2018_PinningDW}. The material-specific structural and magnetic properties lead to variations in DW elastic energies and random pinning characteristics across different systems \cite{Ojha2023_LowCurrentSkyrmion, Jeudy2018_PinningDW, DW_Ojha_ApplPhysA}. Although these differences exist, it has been demonstrated that the creep and depinning dynamics of DWs exhibit universal behavior in ferromagnetic materials \cite{Jeudy2016_UniversalPinning, Kim_Nature_UniversalDynamics, DiazPardo_PRB_2017}.

Recent proposals advocate for using antiferromagnetic (AFM) materials to boost the DW velocity due to the complete suppression of the angular precession \cite{Kim2017_FastDWFerri, Siddiqui_FastDW_Ferri, Gomonay_FastDW_AFM, Panigrahy_PRB_SAFSkyrmionInertia}. However, unlike the ferromagnets, the behaviour of the various dynamical regimes for DW dynamics, i.e. creep, depinning, and flow \cite{Kleeman2007_UniversalDW,Ferre2013_UniDW, DWRelax_Niru_TSF} remains poorly understood for antiferromagnets. Additionally, manipulating magnetic textures and DWs in AFM materials is known to be challenging due to their immunity to external magnetic fields \cite{Song_ReviewAFM_JPD, Mallick_Skyrmion_OAM}. Ferrimagnets, on the other hand, present a significant advantage: they have reduced magnetization and angular momentum like AFMs, but retain control mechanisms similar to ferromagnets due to their non-zero magnetization and near-metallic properties \cite{Stanciu_TAC_Ferri,Eloi_AFM_DW, Berges_PRB_Ferri}. In this context, our approach has been to examine an uncompensated antiferromagnetically coupled bilayer, which behaves as a synthetic ferrimagnetic system to understand the impact of AFM coupling on DW motion and pinning.

Another approach to moving DWs is through current induced spin torques, specifically the classical spin transfer effects in magnetic tunnel junction (MTJ) \cite{Chanthbouala2012_DW} or more recently the spin-orbit torques \cite{Thiaville2012_DMIDW, Parkin2008_DWRacetrack, Beach_ChiralDWMotion, Mallick_PRAppl_0fieldSkyrmion}. Simultaneously enhancing the spin torques and reducing the pinning is essential for achieving efficient current-driven DW motion. To address this, we investigate the  DW dynamics in a transition metal (TM) - rare earth (RE) multilayer, namely Pt$|$Co$|$Tb$|$Al, which forms a synthetic ferrimagnetic heterostructure \cite{Mallick2023_FerriSkyrmion}. By carefully balancing the magnetic moments of the Co and Tb layers, we achieve enhanced DW velocity in the creep regime. The insertion of the Tb layer reduces the anisotropy of Pt$|$Co$|$Al-based multilayers, thereby decreasing the DW pinning energy and increasing the pinning length. By comparing field- and current-driven DW motion, we demonstrate that the efficiency of the spin torque can be accurately estimated in these systems.

\section{\label{sec:methods}Sample description \& experimental methods}

The samples investigated were grown at room temperature on thermally oxidized Si substrates in a high-vacuum sputtering chamber with a base pressure of $5\times10^{-8}$\,mbar. The multilayered samples have the following structure: $||$Ta (5\,nm)$|$Pt (5\,nm)$|$[Pt (3\,nm)$|$Co($t_{\rm Co}$)$|$Tb($t_{\rm Tb}$)$|$Al(3\,nm)]$_{\times N}$$|$Pt(2\,nm), where $t_{\rm Co}=1.0$\,nm for single repetition, i.e. $N=1$ (denoted SL in the following), and 1.3\,nm for $N=5$ (denoted ML), $t_{\rm Tb}=0.25-1.0$\,nm, $||$ indicates the thermally oxidized Si substrate, covered by 280\,nm of SiO$_2$. All the multilayers were grown on top of Ta (5\,nm)$|$Pt (8\,nm) buffers to help stabilizing perpendicular magnetic anisotropy (PMA) in Co and capped with 2\,nm Pt to protect from oxidation. It is worth noting that the thickness of a single atomic layer of Tb is about 0.36\,nm. To facilitate the comparison with a ferromagnetic system, we also prepared a single-layer sample without any Tb layer with the following structure: $||$Ta (5\,nm)$|$Pt (8\,nm)$|$Co(1\,nm)$|$Al(3\,nm)$|$Pt(2\,nm) labeled in the following as Pt$|$Co$|$Al. The growth rates of the layers have been calibrated using x-ray reflectivity measurements. The magnetic moment of the individual layers was quantified by performing x-ray magnetic circular dichroism (XMCD) measurements (on DEIMOS beamline at SOLEIL synchrotron, France \cite{DEIMOS_XMCD}) at 300\,K and 2\,T across the $L_{2,3}$ and $M_{4,5}$ edges of Co and Tb, respectively. The amplitude of the interfacial Dzyaloshinskii–Moriya interaction (i-DMI) in this system was estimated by Brillouin light scattering (BLS) experiments in Damon-Eshbach geometry \cite{Belmeguenai_DMI_BLS}. The field induced dynamics of the DWs was imaged using a Kerr microscope equipped with a micro-coil placed beneath the sample to provide external magnetic field pulses \cite{Jue_DWDynamics_PRB}. For current induced dynamics of the DWs was studied using magnetic force microscopy (MFM) measurements to image sequences of DW displacements after 10-ns long current pulses of varying amplitude. The presence of metallic Tb down to its smallest thickness was confirmed through X-ray photoelectron spectroscopy (XPS) measurements. The stability of the oxygen-sensitive Tb was validated by ensuring consistent data reproducibility over the course of time across multiple measurements.

\section{\label{sec:discuss} Results \& Discussion}
\subsection{\label{sec:magvsTb}Magnetization and Tb thickness}

\newcolumntype{M}[1]{>{\centering\arraybackslash}m{#1}}
\begin{table*}[t]\centering
    \caption{\textbf{Magnetic and depinning parameters at room temperature.} $M_{\rm s}$ is the spontaneous magnetization (deduced from XMCD measurements for SL and from SQUID measurements for ML) calculated using the total Co and Tb volume, $H_{\rm K}$ is the effective anisotropy field, $K_{\rm U}$ and $K_{\rm eff}$ are the effective interfacial and total uniaxial anisotropy energy densities. $T_{\rm d}$, $H_{\rm d}$ and $v(H_{\rm d})$ are depinning parameters as defined in the text. SL and ML indicate the samples with respectively 1 and 5 repetitions.}
    \vspace{6pt}
    \label{TableMagDepin}
	\begin{tabular} {|*{8}{M{2cm}|}}
	\hline
	$t_{\rm Tb}$\,(nm) & $M_{\rm s}$\,(kA\,m$^{-1}$) & $\mu_0H_{\rm K}$\,(mT) & $K_{\rm U}$\,(kJ\,m$^{-3}$) & $K_{\rm eff}$\,(kJ\,m$^{-3}$) & $T_{\rm d}$\,(K) & $\mu_0H_{\rm d}$\,(mT)& $v(H_{\rm d})$\,(m\,s$^{-1}$)\\
		\hline
		0 (SL)   & 1150 $\pm$ 120   & 1800 $\pm$ 90 & 1870 $\pm$ 230    & 1040 $\pm$ 120 &   11000 $\pm$ 1000 & 250 $\pm$ 20 &   100 $\pm$ 20\\
		0.25 (SL)   & 1020 $\pm$ 100   & 730 $\pm$ 40 & 1030 $\pm$ 130    & 370 $\pm$ 40 &   7800 $\pm$ 1000 & 55 $\pm$ 5 &    70 $\pm$ 10\\
		0.50 (SL)   & 820 $\pm$ 80   & 1030 $\pm$ 50  & 850 $\pm$ 100    & 420 $\pm$ 50 &   6800 $\pm$ 1000 & 90 $\pm$ 5 &    80 $\pm$ 10\\
		0.75 (SL)   & 630 $\pm$ 60  & 1390 $\pm$ 70  & 690 $\pm$ 90    & 440 $\pm$ 50 &   8000 $\pm$ 1000 & 170 $\pm$ 10 &    115 $\pm$ 10\\
		1.00 (SL)   & 500 $\pm$ 50  & 1820 $\pm$ 90 & 610 $\pm$ 70    & 450 $\pm$ 50 &   8500 $\pm$ 1000 & 170 $\pm$ 10 &    115 $\pm$ 10\\
        0.25 (ML)   & 990 $\pm$ 80 & 480 $\pm$ 30  & 860 $\pm$ 110    & 240 $\pm$ 30 &   $-$ & $-$ &    $-$\\
        0.50 (ML)   & 930 $\pm$ 50 & 510 $\pm$ 30  & 810 $\pm$ 100    & 240 $\pm$ 30 &   $-$ & $-$ &    $-$\\
        0.75 (ML)   & 620 $\pm$ 60 & 1010 $\pm$ 50  & 690 $\pm$ 90    & 360 $\pm$ 40 &   6500 $\pm$ 1000 & 115 $\pm$ 10 &    130 $\pm$ 10\\
        1.00 (ML)   & 430 $\pm$ 30 & 1370 $\pm$ 70  & 610 $\pm$ 80    & 400 $\pm$ 40 &   7000 $\pm$ 1000 & 120 $\pm$ 10 &    125 $\pm$ 10\\
		\hline
	\end{tabular}
\end{table*}

\begin{figure}[htb]
\centering
\includegraphics[width=8.25cm]{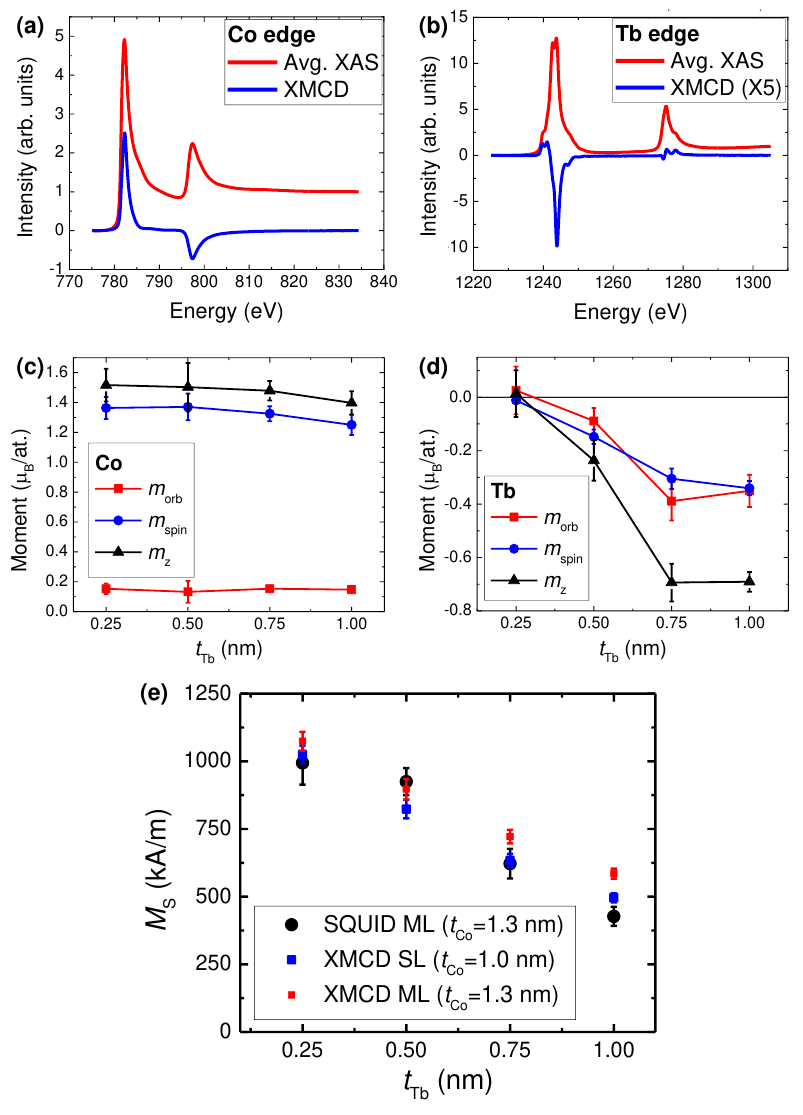}
\caption{\label{fig:1} Background corrected normalized average XAS along with the XMCD spectra along (a) Co and (b) Tb edges. The XMCD signal has been multiplied 5 times. The orbital ($m_{\rm orb}$: red), spin ($m_{\rm spin}$: blue), and total magnetic moments ($m_{z}$: black) corresponding to Co and Tb are presented in (c) and (d), respectively. (e) Comparison between magnetization measured by SQUID magnetometry and deduced from XMCD measurements using Eq.~\ref{Magnetization}.}
\end{figure}

The quantitative estimation of the spin and orbital moments of the individual magnetic elements present in the samples have been obtained by XMCD measurements at 300\,K and $\pm2$\,T at the $L_{2,3}$ and $M_{4,5}$ edges of Co and Tb, respectively. A detailed procedure for the normalization and background correction of the X-ray absorption spectroscopy (XAS) raw data is provided in Appendix A. In Fig.\ref{fig:1}(a) and (b), we show the normalized, background corrected, averaged (left and right circular polarization and positive and negative out-of-plane fields) XAS spectra (red curve) along with the XMCD signal (blue curve) at the Co and Tb edges. The opposite sign of the XMCD spectra for the two atoms confirms the anti-parallel arrangement of the Co and Tb magnetic moments. We quantify the spin ($m_{\rm spin}$) and orbital ($m_{\rm orb}$) atomic magnetic moments for Tb and Co, using the so-called XMCD sum rules \cite{singha_PRB2017, Chen_PRL_1995, Suzuki2023_XMCD-CoTb, SumRule_XMCD_PRB}:

\begin{eqnarray}
    m_{\rm orb}^{\rm Tb}=-\frac{2q}{r}N_{\rm h}^{\rm Tb}\ ;\quad m_{\rm spin}^{\rm Tb}=-2\frac{\langle S_{\rm eff}\rangle }{2+6\frac{\langle T_z\rangle _{\rm free}}{\langle S_z\rangle _{\rm free}}}
    \label{XMCD_Tb}\\
    m_{\rm orb}^{\rm Co}=-\frac{4q}{3r}N_{\rm h}^{\rm Co}\ ;\quad m_{\rm spin}^{\rm Co}=-\frac{6p-4q}{r}N_{\rm h}^{\rm Co}\ .
    \label{XMCD_Co}
\end{eqnarray}

The XAS intensities $I_{+,-}$ are integrated around the $L_3$ energy edge for Co ($M_5$ for Tb), or around both $L_3$ and $L_2$ ($M_4$ and $M_5$), defining the values $p$, $q$ and $r$ for Co or Tb. $+$ or $-$ index corresponds to the sign of the product of the field by the circular polarization, and equivalent XAS are averaged: $p=\int_{L_3 (M_5)}(I_+-I_-){\rm d}E$, $q=\int_{L_3+L_2 (M_5+M_4)}(I_+-I_-){\rm d}E$ and $r=\int_{L_3+L_2 (M_5+M_4)}(I_++I_-){\rm d}E$.
The other parameters are $\langle S_{eff}\rangle =\frac{5p-3q}{r}N_{\rm h}^{\rm Tb}$, ${\langle T_z\rangle_{\rm free}}=-0.243$, ${\langle S_z\rangle_{\rm free}}=-2.943$; $N_{\rm h}^{\rm Tb}$=6, and $N_{\rm h}^{\rm Co}$=2.5, as given by the literature \cite{Suzuki2023_XMCD-CoTb, Streubel_XMCD, Teramura_XMCD}. 

The extracted atomic spin and orbital moments for both Co and Tb layers are plotted as a function of $t_{\rm Tb}$ in Fig.\ref{fig:1}(c) and (d). In Fig. \ref{fig:1}(e), we compare the evolution of the magnetization estimated from XMCD measurements with the one from SQUID measurements. The following equation has been used to obtain the spontaneous magnetization (measured at a field of 2\,T saturating domains, but still below the AFM coupling field) determined from the XMCD moments:

\begin{equation}
    M_{\rm s}
    =\frac{1}{t_{\rm Co}+t_{\rm Tb}}\left( \frac{m_z^{\rm Co}}{V_{\rm Co}}t_{\rm Co}+\frac{m_z^{\rm Tb}}{V_{\rm Tb}}t_{\rm Tb}\right)\quad ,
    \label{Magnetization}
\end{equation}

where $V_{\rm Co}=1.1\times10^{-29}$ m$^3$ and $V_{\rm Tb}=3.2\times10^{-29}$ m$^3$ are the atomic volumes of Co and Tb, and $m_z^{\rm Co}$ and $m_z^{\rm Tb}$ are the magnetic moments of Co and Tb. It can be seen in Fig. \ref{fig:1}(e) that these two independent measurements display a nearly equal evolution of $M_{\rm s}$ with $t_{\rm Tb}$. This validation is essential, as the accuracy of $M_{\rm s}$ is crucial for calculating the effective spin Hall angle ($\theta_{\rm H}$), which will be discussed later in section III-C. As expected, in Fig.\ref{fig:1}(c), the $m_{\rm spin}$ component dominates over $m_{\rm orb}$ in Co due to the significant contribution of the $3d$ orbitals to its magnetism. We note that the total moment of Co for the lowest $t_{\rm Tb}=0.25$\,nm is $\sim1.52\,\mu_{\rm B}$ whereas for a similar Co film without any Tb layer (sample Pt$|$Co$|$Al) is $\sim1.65\,\mu_{\rm B}$. Furthermore, the Co moment decreases from $\sim1.52\,\mu_{\rm B}$ to $\sim1.39\,\mu_{\rm B}$ with the increase of $t_{\rm Tb}=0.25$ to $1$\,nm. The decrease in the Co moment can be understood through the discontinuous environment model \cite{Jaccarino_XMCD_PRL}. According to this model, the magnetic moment of a TM atom is either zero or maximal, based on the number of surrounding transition metal (TM) atoms. As the probability of Tb atoms surrounding a Co atom increases with increasing $t_{\rm Tb}$, the average magnetic moment of Co atoms effectively decreases. This explanation also holds true in the absence of Tb atoms, when Co atoms are only surrounded by Al atoms. In this case, we observe a magnetic moment of $\sim 1.65\,\mu_{\rm B}$, which is slightly lower than the expected $\sim 1.71\,\mu_{\rm B}$ for bulk Co. However, the decrease in Co moment near a Tb atom is stronger than for Al atoms, possibly due to fact that RE and TM elements are extremely prone to alloy formation \cite{Buschow_Physica_RETM_Alloy,Haneswn_JAP_RETM_alloy}. 

Regarding Tb, the magnetic moment increases with $t_{\rm Tb}$. The orbital moment $m_{\rm orb}$ is equivalent to $m_{\rm spin}$ in Tb due to partially filled $4f$ orbitals. Interestingly, the increase of both Tb spin and orbital moments saturates between $t_{\rm Tb} = 0.75$ and $1$\,nm. It is important to note that at room temperature, Tb atoms are supposed to be non-magnetic, in contrast to our observation when they are in contact with Co atoms. The mixing between the two species is localised at the interface in such ferrimagnetic multilayers, which explains the saturation of the Tb magnetic moment beyond $t_{\rm Tb}=0.75$\,nm [Fig.\ref{fig:1}(d)]. This observation is further supported by the relatively low total magnetic moment of Tb in our samples, measuring $\sim 0.6\,\mu_{\rm B}$, compared to the ones reported for TbCo alloys ($\sim 2\,\mu_{\rm B}$) \cite{Suzuki2023_XMCD-CoTb}. As a combination, the total magnetization ($m_z^{\rm Co}+m_z^{\rm Tb}$) of the samples decreases with the increase of $t_{\rm Tb}$.

\begin{figure*}
\centering
\includegraphics[width=18cm]{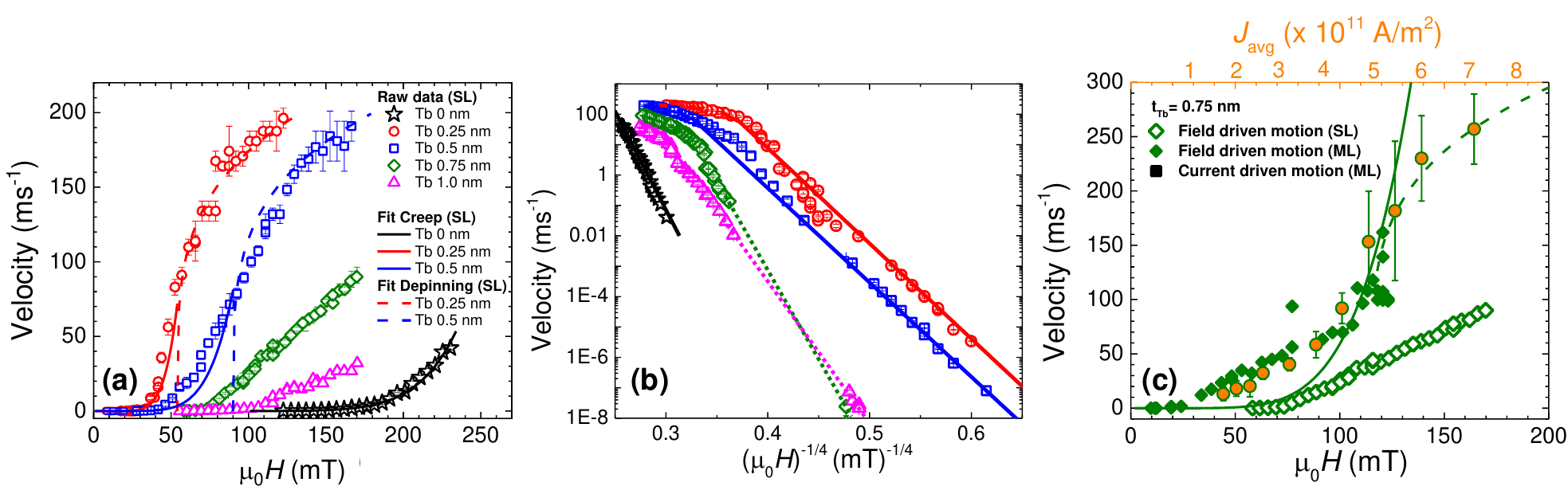}
\caption{\label{fig:2}(a) DW velocity as a function of the external magnetic field for the SL samples with $t_{\rm Tb}=0$\,nm (black stars), $t_{\rm Tb}=0.25$\,nm (red circles), $t_{\rm Tb}=0.5$\,nm (blue squares), $t_{\rm Tb}=0.75$\,nm (green diamonds), and $t_{\rm Tb}=1.0$\,nm (pink triangles). The solid and dashed lines represent the creep (Eq.~\ref{creep}), and depinning (Eq.~\ref{depinning}) fits, respectively. (b) Same curves plotted in semi-log scale vs $(\mu_0H)^{-\frac{1}{4}}$, to evidence the creep regime. (c) Comparison between field-induced DW motion in a single layer (SL: open diamond symbol), multilayer (ML: filled diamond symbol) and current-induced DW motion in a multilayer (ML: green circles filled with orange) for $t_{\rm Tb}=0.75$\,nm. The current density scale bar for the ML sample is given as the top axis in orange color. For the ML, the best superimposition between current- and field-induced velocity curves was obtained using a multiplicative conversion factor $\frac{\mu_0H}{J_{\rm avg}}=27.5\times10^{-11}$\,mT\,A$^{-1}$m$^2$ for transforming the current density into field. The solid and dashed lines respectively correspond to Eq.~\ref{creep} and Eq.~\ref{depinning}; they highlight the crossover between creep and depinning.}
\end{figure*}

A detailed summary of the various micromagnetic parameters of the samples (extracted from M-H loops measured using AGFM, and SQUID magnetometry), viz. spontaneous magnetization ($M_{\rm s}$), anisotropy field ($H_{\rm K}$), uniaxial anisotropy ($K_{\rm U}$), and effective anisotropy ($K_{\rm eff}$) is provided in Table \ref{TableMagDepin}, where $K_{\rm U}=K_{\rm eff}+\frac{1}{2}\mu_0{M_{\rm s}}^2$. We note that, as explained by the XMCD data, there is a continuous decrease in $M_{\rm s}$ with increasing $t_{\rm Tb}$ owing to the AFM coupling. For $H_{\rm K}$, $K_{\rm U}$ and $K_{\rm eff}$, we observe a sharp decrease from $t_{\rm Tb}=0$ to 0.25\,nm. This suggests that the interface anisotropy of Co$|$Al is larger than that of Co$|$Tb, due to the detailed electronic reconstruction at the Co$|$Al interface \cite{Sachin_NanoLetter,Gamberdella_PRB_OrbitalAnisotropy,sachin_future}. It should be noticed that there is a monotonic decrease (increase) of $M_{\rm s}$ ($H_{\rm K}$) in the thickness range of $t_{\rm Tb}$ from 0.25 to 1\,nm due to the increase in the magnetic compensation while maintaining the same exchange coupling. However, the change in anisotropy energy is less pronounced since it depends on the product of $M_{\rm s}$ and $H_{\rm K}$. Assuming the exchanges (symmetric $A$ and antisymmetric $D$) constant for all the samples, this is expected to result in a strong decrease of the DW energy ($\sigma=4\sqrt{AK_{\rm eff}}$) and increase of DW width parameter ($\Delta=\sqrt{\frac{A}{K_{\rm eff}}}$) between the samples with $t_{\rm Tb}=0$ and $0.25$\,nm. However, there should not be a substantial change of these two parameters between the samples with various $t_{\rm Tb}$. The evolution of these parameters with $t_{\rm Tb}$ will be discussed in details in section III-D. 

Magnetization dynamics was explored by BLS for the lower Tb thicknesses (0.25 and 0.5 nm), for the single layer (SL) and multi layer (ML) series (see Appendix B). Indeed the linewidth of the peaks is seen to increase markedly as Tb thickness increased. In addition, this linewidth is also seen to decrease when comparing ML to SL samples. The i-DMI amplitude has been extracted from the BLS measurements. We find respectively $D=-1.3\pm0.2$ and $-0.49\pm0.02$\,mJ\,m$^{-2}$ for the SL and ML samples with $t_{\rm Tb}=0.25$\,nm and$-0.3\pm0.07$\,mJ\,m$^{-2}$ for the ML sample with $t_{\rm Tb}=0.5$\,nm. The thicker samples exhibit excessively broad BLS peak linewidths, making i-DMI measurements difficult. Using the values of Table 1, this leads to DMI effective fields $\mu_0 H_{DMI}= 90 (65)$ mT for the ML samples with $t_{\rm Tb}=0.25 (0.5)$\,nm. As the computed Néel wall demagnetizing fields for these samples are 430 (470) mT, respectively, their domain walls are intermediate between achiral Bloch and chiral Néel walls ($\cos \Phi= 0.33 (0.22)$).

\subsection{\label{sec:fielddyna}Field-induced DW dynamics}

The impact of AFM coupling on the dynamical behavior of the DW has been investigated through two series of measurements where they are driven either by out-of-plane magnetic field or by current pulses. First we examine the field induced DW motion. The velocity curves for the different samples displayed in Fig.\ref{fig:2} (a) and (b) for the SL samples with various $t_{\rm Tb}$ reveal the pinning dependent regimes for the DW dynamics. For clarity, we will use consistent color schemes throughout the article to represent data for samples with different $t_{\rm Tb}$. It is important to note that for all the curves, we set the upper limit of the explored magnetic field range based on the field at which multiple domain nucleation events are observed during the magnetic field pulse, as this interferes with velocity measurements. For the same reason, measurements could also not be conducted for ML samples with $t_{\rm Tb}< 0.75$\,nm. As observed, changing $t_{\rm Tb}$ strongly alters the shape of velocity curves and the field range over which significant velocities are observed. Specifically, increasing $t_{\rm Tb}$ shifts the curves towards larger magnetic fields except for the film without Tb, which is observed at the highest field. To identify the different dynamical regimes and quantitatively analyze the evolution of DW pinning with $t_{\rm Tb}$, the velocity curves have been compared to (i) the self-consistent description of the creep regime (dashed lines, $H \leq H_{\rm d}$) 
\begin{equation}
    v_{\rm creep}(H,T)=v(H_{\rm d},T)\exp\left(-\frac{\Delta E}{k_{\rm B}T}\right)\quad,
    \label{creep}
\end{equation}
where $k_{\rm B}T$ is the thermal activation energy, $\Delta E=k_{\rm B}T_{\rm d}\left[\left(\frac{H}{H_{\rm d}}\right)^{-\mu}-1\right]$ the effective pinning barrier, and (ii) the depinning regime in non-thermally activated limit (solid lines, $ H \gtrsim H_{\rm d}$), given by
\begin{equation}
v_{\rm dep}(H,T)=\frac{v(H_{\rm d},T)}{x_0}\left(\frac{T_{\rm d}}{T}\right)^\Psi\left(\frac{H-H_{\rm d}}{H_{\rm d}}\right)^\beta\quad ,
\label{depinning}
\end{equation}
as developed for ferromagnetic ultrathin films~\cite{Jeudy2016_UniversalPinning, Jeudy2018_PinningDW, DiazPardo_PRB_2017}. In Eqs.~\ref{creep} and~\ref{depinning}, $\mu=0.25, \beta=0.25, \Psi=0.15$, and $x_0=0.65$ are the universal parameters of the quenched Edwards Wilkinson universality class for thin films and short range pinning interactions. The depinning field $H_{\rm d}$, temperature $T_{\rm d}$, and velocity $v(H_{\rm d})$ are material dependent parameters reflecting the interaction of the DW with disorder. Eqs.~\ref{creep}-~\ref{depinning} describe the dynamics resulting from intricate and cooperative processes shaped by multiple influential factors, namely the interplay between the external driving field, DW elasticity, thermal fluctuations, and inherent disorder of the sample.

As observed in Fig.~\ref{fig:2}(a) and (b), the experimental velocities for the different samples fit rather well with the expectations (plain line) from the creep law (Eq.~\ref{creep}) in the lowest field drive range. Note that the curves for $t_{\rm Tb}=0.25$ and $0.50$\,nm present (red and blue in Fig.~\ref{fig:2}(a)) a change between positive and negative curvature compatible with a crossover between creep and depinning regimes. For $t_{\rm Tb}=0$\,nm, the negative curvature is not observed, indicating that the depinning regime has not been reached. As $t_{\rm Tb}$ increases from $0.50$ to  $0.75$\,nm, we find that the fit with creep law just below the depinning field is not very good. The linear variation of the velocity observed for $t_{\rm Tb}=0.75$\,nm and $t_{\rm Tb}=1.0$\,nm is not typically seen for ferromagnetic films, suggesting that it may be characteristic of DW dynamics in AFM systems. A possible explanation could be a divergence of the avalanche sizes close to depinning. However, the absence of quantitative agreement with the prediction of Ref.~\cite{CaballeroPRE2018_DWCreep} leaves the origin of this behavior as an open question which goes beyond the scope of this work. The depinning parameters ($H_{\rm d}$, $v(H_{\rm d})$, and $T_{\rm d}$) reported in Table~\ref{TableMagDepin} for $t_{\rm Tb}=0.25$ and $=0.50$\,nm were determined by the fits of Eqs.~\ref{creep}-\ref{depinning}. For the other $t_{\rm Tb}$, we assume that the maximum measured velocity is the one close to the depinning transition: the maximum measured field and velocity correspond to $H_{\rm d}$ and $v(H_{\rm d})$, respectively. The depinning temperature is then defined by the best fit of the creep regime (see the dotted lines in Fig. 2(b)).

\begin{figure}[hbt]
\centering
\includegraphics[width=7cm]{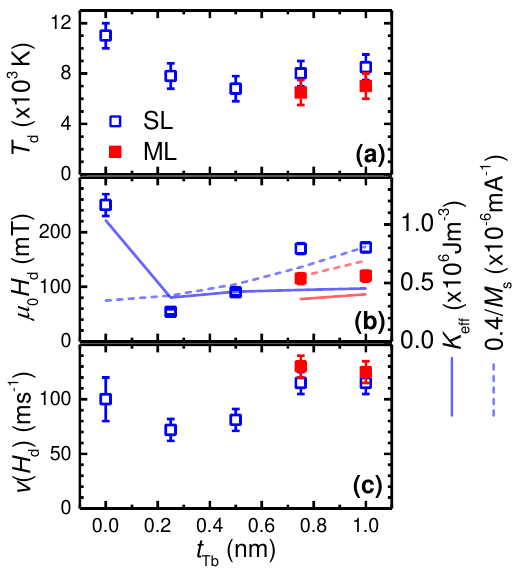}
\caption{\label{fig:depinning}Depinning parameters extracted from the fit of the field-induced DW motion: (a) $T_{\rm d}$, (b) $H_{\rm d}$, and  (c) $v(H_{\rm d})$ (square symbols). In panel (b), $K_{\rm eff}$ is traced as a plain lines to demonstrate the how the anisotropy jump as the first Tb layer is inserted correlates with $H_{\rm d}$. After the initial jump corresponding to the insertion of Tb, $H_{\rm d}$ correlates with $M_{\rm s}^{-1}$ displayed as a dashed line.}
\end{figure}

As shown in Fig.~\ref{fig:depinning}, the depinning field $H_{\rm d}$ displays a non-monotonous variation. We find that $H_{\rm d}$ presents a strong reduction (by a factor 5) between 0 and 0.25\,nm of $t_{\rm Tb}$ and then increases strongly (by more than a factor 3) between $t_{\rm Tb}=$0.25 and 1.0\,nm. The strong reduction of $H_{\rm d}$ between $t_{\rm Tb}=0$ and 0.25\,nm coincides with the sharp increase of the anisotropy. Then by assuming a constant pinning force, as the driving force is proportional to $\mu_0 M_{\rm s} H$, the decrease of $M_{\rm s}$ with increasing $t_{\rm Tb}$ between 0.25 and 1.0\,nm is expected to increase the depinning field $H_{\rm d}$. In Fig.~\ref{fig:depinning}(b), $K_{\rm eff}$ is traced as solid line to demonstrate how the anisotropy jump as the first Tb layer is inserted correlates with $H_{\rm d}$. After the initial jump corresponding to the insertion of Tb, $H_{\rm d}$ correlates with $M_{\rm s}^{\rm -1}$ as displayed by a dotted line.

For the ML samples, the values of $H_{\rm d}$ are smaller than those for the SL samples. This is attributed to a reduction in anisotropy caused by the incremental increase in Co thickness and the influence of dipolar coupling within the ML structure. It also corresponds to the evolution of the BLS peaks linewiths. For $T_{\rm d}$, the variations are also non-monotonous: a pronounced decrease is observed in the SL samples when the first layer of Tb is inserted (from $t_{\rm Tb}=0$ to $0.25$\,nm). This confirms that the DW pinning is significantly reduced by the introduction of Tb layer in the system. Henceforth, the change in $T_{\rm d}$ with further Tb insertion is less significant. The continuous increase observed is still correlated with the rise of $H_{\rm d}$. As the depinning velocity $v_T=v(H_{\rm d},T)/x_0$ depends on $H_{\rm d}$ and on the shape of the velocity curves, it is often misleading to comment on the trend of the velocity. Nevertheless, we observe an increase in $v_T$ with $t_{\rm Tb}$ as well as from the SL to the ML samples. In summary, we find that (i) the strong reduction of the film anisotropy with $t_{\rm Tb}=0.25$\,nm coincides with a strong reduction of $H_{\rm d}$, (ii) the increase of $t_{\rm Tb}$ reduces $M_{\rm s}$, due to AFM coupling between Co and Tb, which enhances $H_{\rm d}$, and (iii) the behaviors are quantitatively similar for SL and ML samples.

\subsection{\label{sec:currentdyna}Current-induced DW dynamics}
In Fig. \ref{fig:2} (c), we present the velocity curves obtained for magnetic field driven DW motion in SL (open diamond), in ML (plain diamond) and for current driven DW motion in ML (open circle filled with orange) with a fixed Tb thickness ($t_{\rm Tb}$=0.75\,nm). As observed, the linear variation of the velocity in the low driving range found in SLs is also observed for the ML sample. The apparent shift of about $50$\,mT is partly due to the larger $M_{\rm s}$ in the ML sample. To compare between current and field driven DW motion, we used a single multiplicative conversion factor to rescale the current density ($J_{\rm avg}$). We observe that the two curves are consistent with a crossover between creep and depinning, as indicated by their comparison with the predictions of Eqs.~\ref{creep}-~\ref{depinning}. The overlap of the curves suggests an equivalence between SOT and a perpendicular magnetic field, as predicted for free DW motion~\cite{Thiaville2012_DMIDW} in the limit $J << J_D$ (which is justified here as one computes $J_D= 120 \times 10^{11}$ A/m$^2$ for the ML with $t_{Tb}= 0.25$ nm, assuming a damping constant $\alpha= 0.1$), and neglecting the FL torque (see App. C for justification).

\begin{figure}[htb]
\centering
\includegraphics[width=7cm]{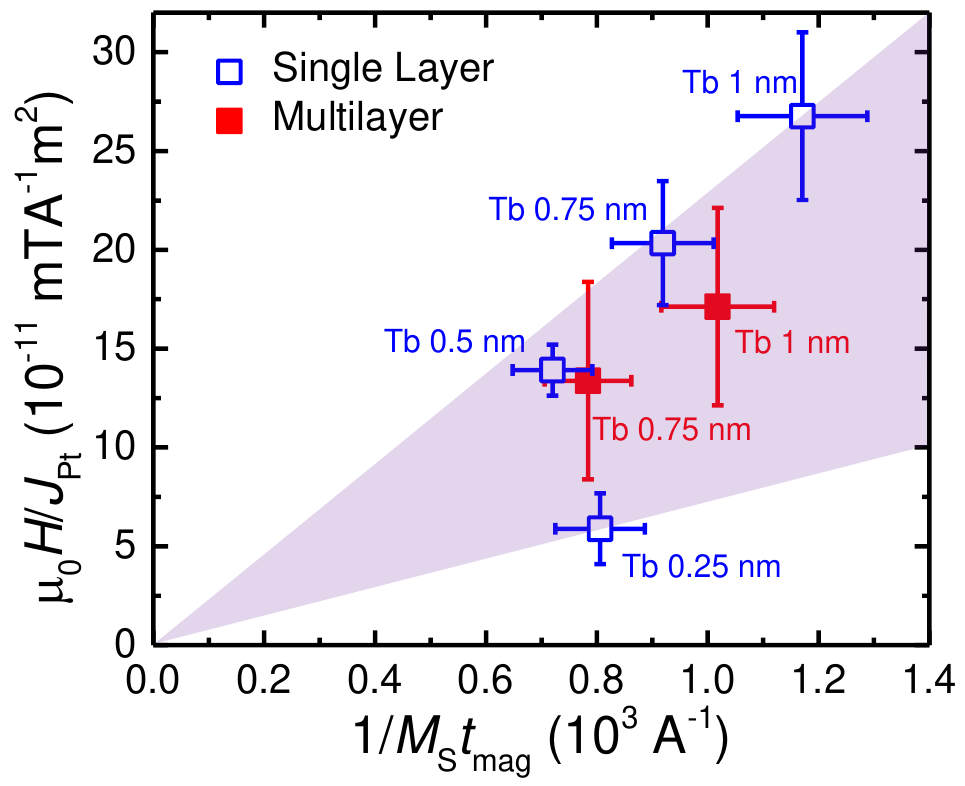}
\caption{\label{fig:SOT} Comparison of conversion factor between field and current driven DW motion for different $t_{\rm Tb}$, plotted as a function of $1/M_{\rm s}t_{\rm mag}$. The current density has been calculated only through the Pt layers.  Within the shaded region, from Eq.~\ref{conversion} and assuming that $\cos\Phi=1$, we estimate $\theta_{\rm H}=0.265\pm0.125$, attributing the primary sources of SOT to the Pt$|$Co interfaces.}
\end{figure}

To gain further insight into the contribution of AFM coupling on DW driven by SOT, the conversion factor ($\frac{\mu_0H}{J_{\rm avg}}$) was estimated for all the samples studied. It is important to highlight that the field-driven DW motion experiments were performed on all SL and some ML samples, as detailed in the previous section. However, current-induced DW motion experiments were only conducted on the ML samples. Therefore, the \textit{SL} series may provide another comparison between field-induced and current-induced DW motion. In appendix C (see Fig. \ref{fig:7}), we show the conversion factor plotted as a function of $\frac{1}{M_{\rm s}t_{\rm mag}}$ where the current density $J_{\rm avg}$ is calculated as it was uniform through all the layers within the heterostructure and $t_{\rm mag}$ is the magnetic layers (Co+Tb) thickness that was used to calculate $M_{\rm s}$. A monotonous increase in the efficiency as a function of $\frac{1}{M_{\rm s}t_{\rm mag}}$ is found by virtue of the AFM coupling in both the SL and ML samples. The increasing trend of the conversion factor aligns well with the modified $q-\Phi$ model, as explained by the Thiele force equation approach \cite{Thiaville2012_DMIDW}:
\begin{equation}
    \frac{\mu_0H}{J}=\frac{h}{8e} \frac{\theta_{\rm H} \cos\Phi}{M_{\rm s}t_{\rm mag}} \quad ,
    \label{conversion}
\end{equation}
where $h$ is the Planck constant, $\theta_{\rm H}$ is the effective spin Hall angle, and $|\cos\Phi|=1$ for pure transversal Néel walls ($\Phi$ is the angle between the current density and the in-plane magnetization direction in the DW).

An accurate estimation of $\theta_{\rm H}$ requires to disentangle the actual proportion of current flowing through different layers within the ML structure. Previous research has shown that the predominant source of SOT arises from the Pt$|$Co interface \cite{Sachin_NanoLetter}. Therefore, the conversion factor has been calculated in terms of current flowing through the Pt layers ($J_{\rm Pt}$) only. The current density through the individual layers has been estimated by measuring their resistivity in four-point geometries. The resistivity $\rho_i$ of Ta, Co, and Pt films of similar thicknesses are respectively 169, 30 and 24\,$\mu\Omega$\,cm. The current density through the Tb and Al layers has been neglected due to the high resistivity of these materials at the thicknesses that we experimentally measured. To estimate the current density in the Pt layers, we used~\cite{Sachin_NanoLetter}:
\begin{equation}
    J_{\rm Pt}\approx\frac{ t_{\rm tot} }{ t_{\rm Pt}+\frac{\rho_{\rm Pt}}{\rho_{\rm Ta}}t_{\rm Ta}+\frac{\rho_{\rm Pt}}{\rho_{\rm Co}}t_{\rm Co}}J_{\rm avg} \quad ,
\end{equation}
where $t_{\rm tot}$ is the total thickness of the heterostructure. In Fig. \ref{fig:SOT}, we present the rescaled conversion factor ($\frac{\mu_0H}{J_{\rm Pt}}$) as a function of $\frac{1}{M_{\rm s}t_{\rm mag}}$. The data in Fig. \ref{fig:SOT} can be fitted using Eq.~\ref{conversion}, substituting $J$ by $J_{\rm Pt}$. The data points show some dispersion and do not strictly follow a linear trend. To account for this variability, we have shaded a region in Fig. \ref{fig:SOT} to represent the fitting range, capturing the dispersion of the individual points. Within this region, we estimate $\theta_{\rm H}=0.265\pm0.125$, attributing the primary sources of SOT to the Pt$|$Co interfaces. Additionally, the effective spin Hall angle $\theta_{\rm H}=0.086\pm0.006$ was measured in the SL samples using the second harmonic Hall measurement technique (see Appendix D for details), in which the bottom 8\,nm Pt is again considered being the only source of SOT. There are several potential reasons for the mismatched $\theta_{\rm H}$ values obtained from the two different measurement methods. One reason is that the ML and SL samples used for current- and field-induced dynamic measurements were not identical, as previously noted. However, as shown in Fig.\ref{fig:SOT}, the SL points are in the same range of values as the ML points. Additionally, the Joule heating produced by the current pulse is expected to heat the track \cite{TrackHeating, JouleHeating}. For a typical current density $J=3.5\times10^{11}$ A$\rm m^{-2}$, and pulse duration $t=20$ ns, the heat is expected to be confined into the $\rm SiO_{2}$ layer just below the track (see Appendix E for details). The estimated temperature rise $\Delta T(10\rm ns) \sim 220$K is rather large. On the one hand, it is expected to reduce significantly the DW width parameter $\Delta$ and to produce highly non-linear velocity enhancement \cite{JouleHeating}. In addition, the associated decrease of the spontaneous magnetization directly increases (see Eq.~\ref{conversion}) the effective field due to the SOT. Analyzing the contribution of Joule heating to DW velocity for each value of current density and accounting for the dynamical regimes (creep, depinning, and flow) followed by DW is beyond the scope of this work. On the other hand, another source of error arises from the calculation of $J_{\rm Pt}$, due to the intrinsic complexities and assumptions involved in this method. Taking these factors into account, we believe that the spin Hall angle estimated from the DW dynamic measurements is in reasonable agreement  with the results from the second Harmonic Hall measurements. This alternative method for estimating the spin Hall angle in multilayers is interesting, as traditional techniques like second harmonic Hall or inverse spin Hall effect measurements, along with their analyses, can be cumbersome.

\subsection{\label{sec:disorder}Influence of DW disorder interaction}

\begin{figure}[htb]
\centering
\includegraphics[width=8.75cm]{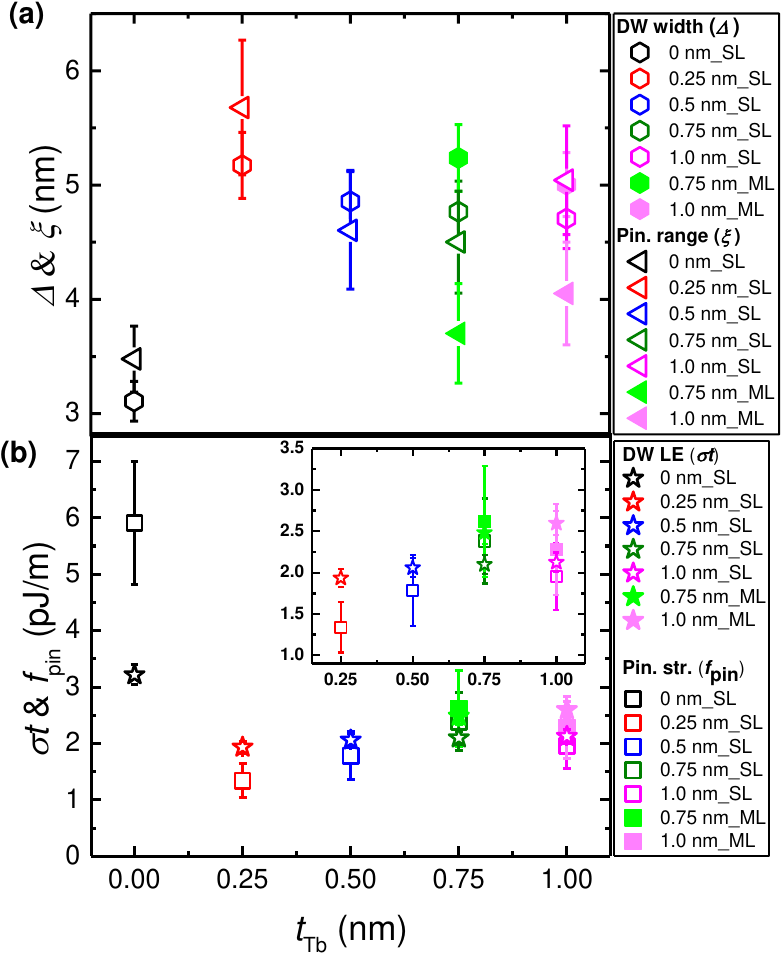}%
\caption{\label{fig:PinPara}Pinning parameters of the samples: (a) variation of DW width parameter ($\Delta$), and pinning length ($\xi$) as a function of $t_{\rm Tb}$. (b) variation of pinning strength ($f_{\rm pin}$) and DW line energy (DW LE: $\sigma t$) as a function of $t_{\rm Tb}$. The filled and open data points represent the data for the SL and ML samples, respectively.}
\end{figure}

To discuss the evolution of the interaction between DWs and disorder with the Tb thickness in SL and ML, we apply the scaling model developed in Ref.~\cite{gehanne_prr_2020}. This model relates the depinning parameters ($H_{\rm d}$ and $T_{\rm d}$) to the characteristic length scale  $\xi$ and the force of the DW disorder interaction $f_{\rm pin}$ :
\begin{eqnarray}
	\label{eq: 3_xi}
	\xi \sim \left[ (k_{\rm B}T_{\rm d})^2/(2\mu_0 H_{\rm d}M_{\rm s}\sigma t_{\rm mag}^2)\right]^{1/3}\quad ,\\
	\label{eq: 2_fpin}
	f_{\rm pin} \sim \frac{b}{\xi}\sqrt{2\mu_0 H_{\rm d}M_{\rm s}t_{\rm mag}k_{\rm B}T_{\rm d}}\quad ,
	\label{eq:xi_fpin}%
\end{eqnarray}
where the $\sigma$ is the DW energy density with $A=10$\,pJ/m the exchange stiffness, $k_{\rm B}$ is the Boltzmann constant, and $b$ is a characteristic distance between pinning sites. From the data of table I, we obtain the variations of $\xi$ and $f_{\rm pin}$ as a function of $t_{\rm Tb}$ shown in Fig.~\ref{fig:PinPara}(a). We first compare the variation of DW width parameter $\Delta$ and $\xi$. For the DW width $\Delta$, we find a strong increase between 0 and 0.25nm of Tb, associated to the decrease of $K_{\rm eff}$ by the factor 3 (see Table I). The values obtained for the MLs remains close to those obtained for the SLs. As for the pinning range $\xi$, we observe a good scaling $\Delta \sim \xi$ for all $t_{\rm Tb}$. We used a single scaling factor (=1.5) for $\xi$ to obtain the best superimposition between with the values of $\Delta$. This result is compatible with a typical distance between pinning centers $b$ smaller than the DW thickness ($\approx\pi\Delta$), which thus corresponds to the DW pinning range~\cite{gehanne_prr_2020,Balan_APL_2023}. 

In Fig.~\ref{fig:PinPara}(b), we compare the variation of the pinning strength $f_{\rm pin}$ and the DW line energy $\sigma t$ as a function of $t_{\rm Tb}$. For the DW energy $\sigma t$ , we find a strong decrease between $t_{\rm Tb}=0$ and $0.25$\,nm associated to the jump of $K_{\rm eff}$. For the calculation of $f_{\rm pin}$, we choose arbitrarily $b= 1$\,nm and use a single scaling factor ($= 7.5$) to obtain the best superposition with $\sigma t$. Between $t_{\rm Tb}= 0.25$ and  1\,nm, variations of $f_{\rm pin}$ scale rather well with those of $\sigma t$, in agreement with the findings of Refs. ~\cite{gehanne_prr_2020,Balan_APL_2023}, which were obtained with a fixed disorder. Therefore, the good scaling strongly suggests that the elaboration method is sufficiently reliable to produce a reproducible pinning disorder. Note also that $f_{\rm pin}$ and $\xi$ present opposite curvature for their variations  with $t_{\rm Tb}$. Such behavior is expected since for $\pi\Delta>b$, the interaction of DW is smoothed over the DW thickness. For a fixed disorder, reducing $\Delta$ is expected to enhance the interaction strength. Moreover, it should be noted that without Tb layer, a bad scaling between $f_{\rm pin}$ and $\sigma t$ is observed [Fig.~\ref{fig:PinPara}(b)]. Therefore, the changes of effective anisotropy and magnetization with $t_{\rm Tb}$ at the origin of the scaling between $\xi$ and $\Delta$ are not sufficient to explain the variation of DW pinning. The high value of $f_{\rm pin}$ suggests that the pinning strength is larger for the Co$|$Al interface than for the Co$|$Tb interface. This effect may be both associated to enhanced fluctuations of effective anisotropy and to a reduced correlation of pinning disorder.
\section{Conclusion}
In conclusion, we have investigated the impact of incorporating a Tb layer in Pt$|$Co$|$Al synthetic ferrimagnetic multilayers on field- and current-driven DW motion. We show that inserting a Tb layer between the Co and Al layers reduces anisotropy, thereby weakening the interaction strength between the DW and disorder. This antiferromagnetic coupling leads to a decrease in pinning strength, resulting in faster DW motion. Additionally, we introduce a novel method to quantify effective spin torques in these multilayers by analyzing the depinning transitions. Our work not only enhances our understanding of DW mobility in ferrimagnetic multilayers but also offers strategies for optimizing spin-orbit torques and pinning landscapes. We believe this research is a crucial step towards the realization of DW logic and memory devices using antiferromagnetic or ferrimagnetic materials, bridging the gap between theoretical understanding and practical implementation.

\begin{acknowledgments}
We acknowledge Dr. Yanis Sassi for sharing data of Pt$|$Co$|$Al multilayers, Dr. Sachin Krishnia and Dr. Henri Jaffrès for fruitful discussion. The XMCD experiments were performed on the DEIMOS beamline at SOLEIL Synchrotron, France and the authors are grateful to the SOLEIL staff for smoothly running the facility. This work has been financially supported by a government grant managed by the French ANR as part of the “Investissements d’Avenir” program (Labex NanoSaclay, SPiCY ANR-10-LABX-0035) and as a part of the France 2030
investment plan from PEPR SPIN CHIREX ANR-22-EXSP-0002 and SPINMAT  ANR-22-EXSP-0008. We also ackowledge the support from the Horizon2020 Framework Program of the European Commission, under FETProactive Grant agreement No. 824123 (SKYTOP) (H2020 FET proactive 824123).
\end{acknowledgments}

\appendix

\section{XMCD data treatment}
\begin{figure}[htb]
\centering
\includegraphics[width=8.75cm]{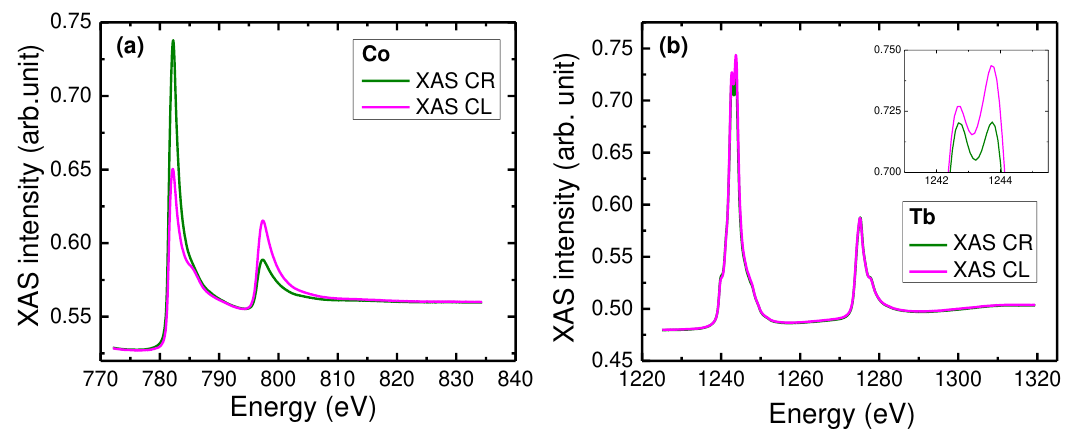}%
\caption{\label{fig:6}Average XAS spectra for right-(CR: green) and left-circular(CL: pink) polarized light for (a) Co and (b) Tb under an out-of-plane external field of 2\,T.}
\end{figure}

In Fig.\ref{fig:6} (a) and (b), we show the raw XAS spectra acquired with an external out-of-plane magnetic field of 2\,T with right-circular (CR: green) and left-circular (CL: pink) polarized light for Co and Tb, respectively. The displayed XAS spectra for each polarization are averages of several summed XAS measurements. The remaining XAS background is removed to determine $r$ by subtracting the linear interpolation of the off-peak intervals from the raw spectra for each polarization \cite{SumRule_XMCD_PRB}. The data has been normalized at the pre-edge to coincide with 0, and further renormalized at the post-edge to 1. The normalized average XAS spectra ($\frac{I_++I_-}{2}$) are shown in Fig.\ref{fig:1} (a) and (b). The XMCD signal is obtained by subtracting the normalized CL spectra from the normalized CR spectra (for ``positive'' field). Note that the error in XMCD measurements predominantly arises during total electron yield, due to a distortion of the XAS from saturation effects \cite{XMCD_error}. For the error bars, we have used the difference in values obtained from various spectra in the same configuration or with reversed magnetic field.

\section{BLS Measurements}
\begin{figure}[htb]
\centering
\includegraphics[width=8.5cm]{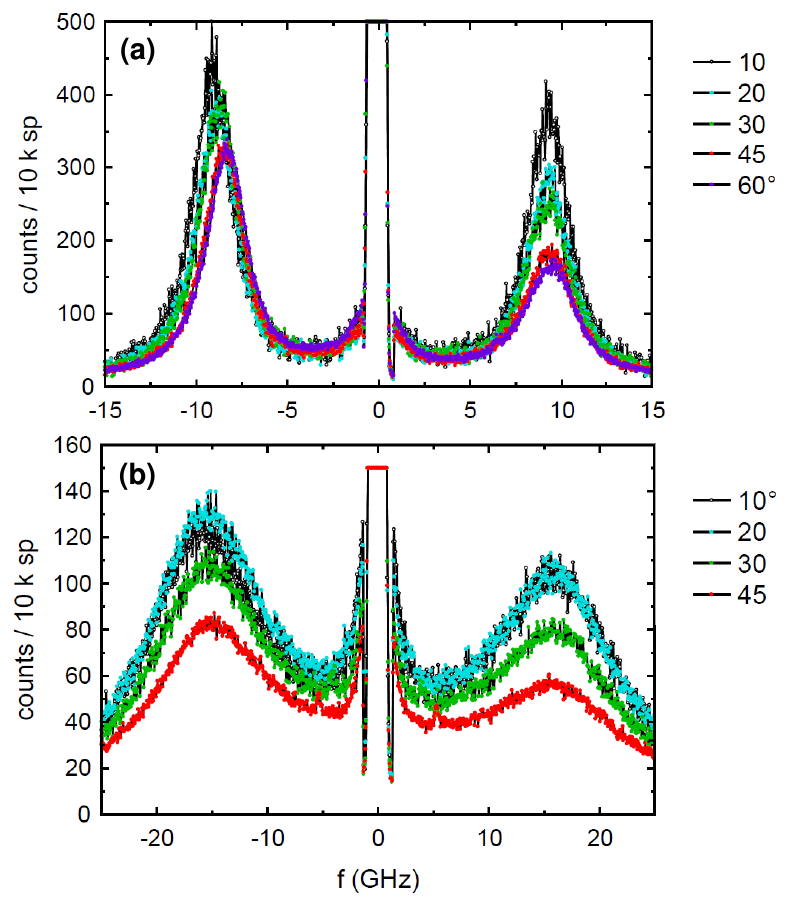}%
\caption{\label{fig:BLS}BLS spectra for the ML samples with (a) $t_{Tb}=0.25$ nm and (b) $t_{Tb}=0.5$ nm as a function of various applied magnetic fields.}
\end{figure}
In order to estimate the interfacial DMI in the heterostructure, BLS measurements have been performed
in Damon-Eschbach geometry on the ML samples with lower Tb thicknesses of $t_{Tb}=0.25$ and $0.5$ nm. It should be noted that the linewidth of the peaks was observed to increase markedly as Tb thickness increased as seen in Fig. \ref{fig:BLS}. In addition, this linewidth was also seen to decrease when comparing ML to SL samples (data not shown here). The DMI can be extracted using $\Delta f=f_S(-k)-f_{AS}(k)=\frac{2\gamma}{\pi M_s}D_{eff}k_{SW}$. We find respectively $D=-1.3\pm0.2$ and $-0.49\pm0.02$\,mJ\,m$^{-2}$ for the SL and ML samples with $t_{\rm Tb}=0.25$\,nm and $-0.3\pm0.07$\,mJ\,m$^{-2}$ for the ML sample with $t_{\rm Tb}=0.5$\,nm.

\section{Comparison of field- and current-induced motion}
\begin{figure}[htb]
\centering
\includegraphics[width=7.5cm]{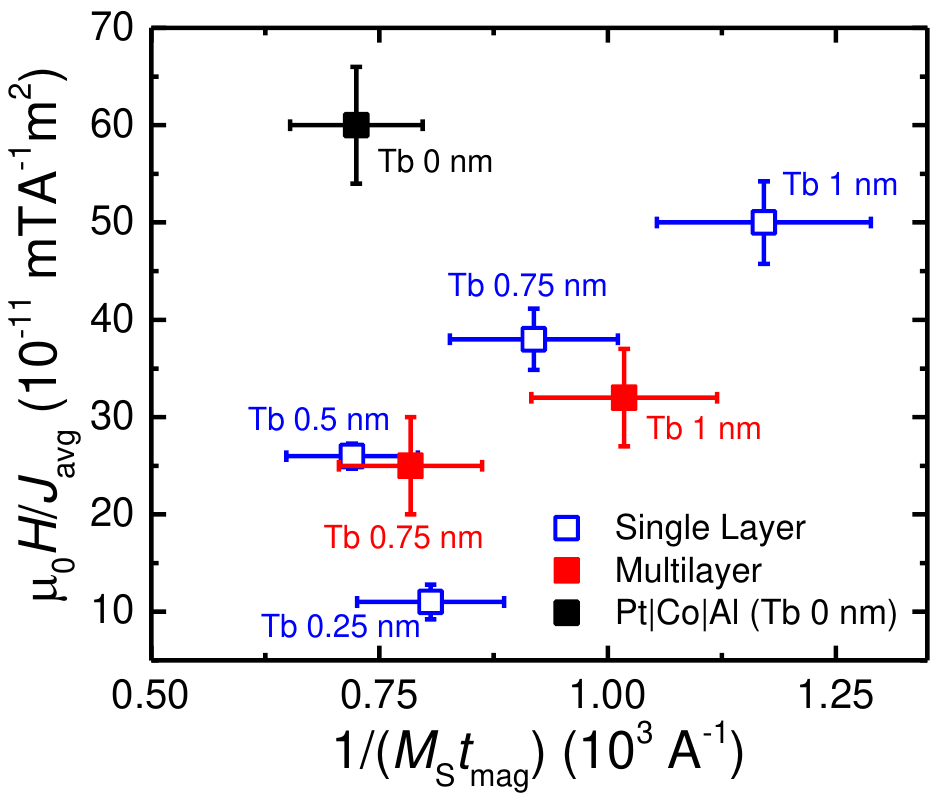}%
\caption{\label{fig:7}Comparison of conversion factor between field and current driven DW motion for different $t_{\rm Tb}$, plotted as a function of $1/M_{\rm s}t_{\rm mag}$. The current density is calculated as if it was uniform throughout the sample thickness.}
\end{figure}
In Fig. \ref{fig:7}, we show the conversion factor ($\frac{\mu_0H}{J_{\rm avg}}$) plotted as a function of $\frac{1}{M_{\rm s}t_{\rm mag}}$ where the current density $J_{\rm avg}$ is calculated as if it was uniform through all the layers within the heterostructure. To ensure accurate selection of $M_{\rm s}$, the values have been extracted from XMCD and SQUID measurements for the SL and ML samples, respectively. A monotonous increase in the efficiency has been observed in all the samples. It is important to note that current-induced DW motion experiments have been conducted exclusively on ML samples, while field-induced DW motion experiments have been performed on both SL and ML samples. To compare the results from these two types of experiments, the conversion factor has been used as a scaling factor to superimpose the data. We note that the conversion factor is rather small for the SL sample with $t_{\rm Tb}=0.25$\,nm. This is primarily due to the strong drop in anisotropy with the insertion of Tb layer into the heterostructure. Additionally, the Tb layer thickness is less than a monolayer for this sample, leaving the system not well optimized for efficient SOT driven DW motion. 

\section{Second harmonic Hall measurements}
\begin{figure}[htb]
\centering
\includegraphics[width=9cm]{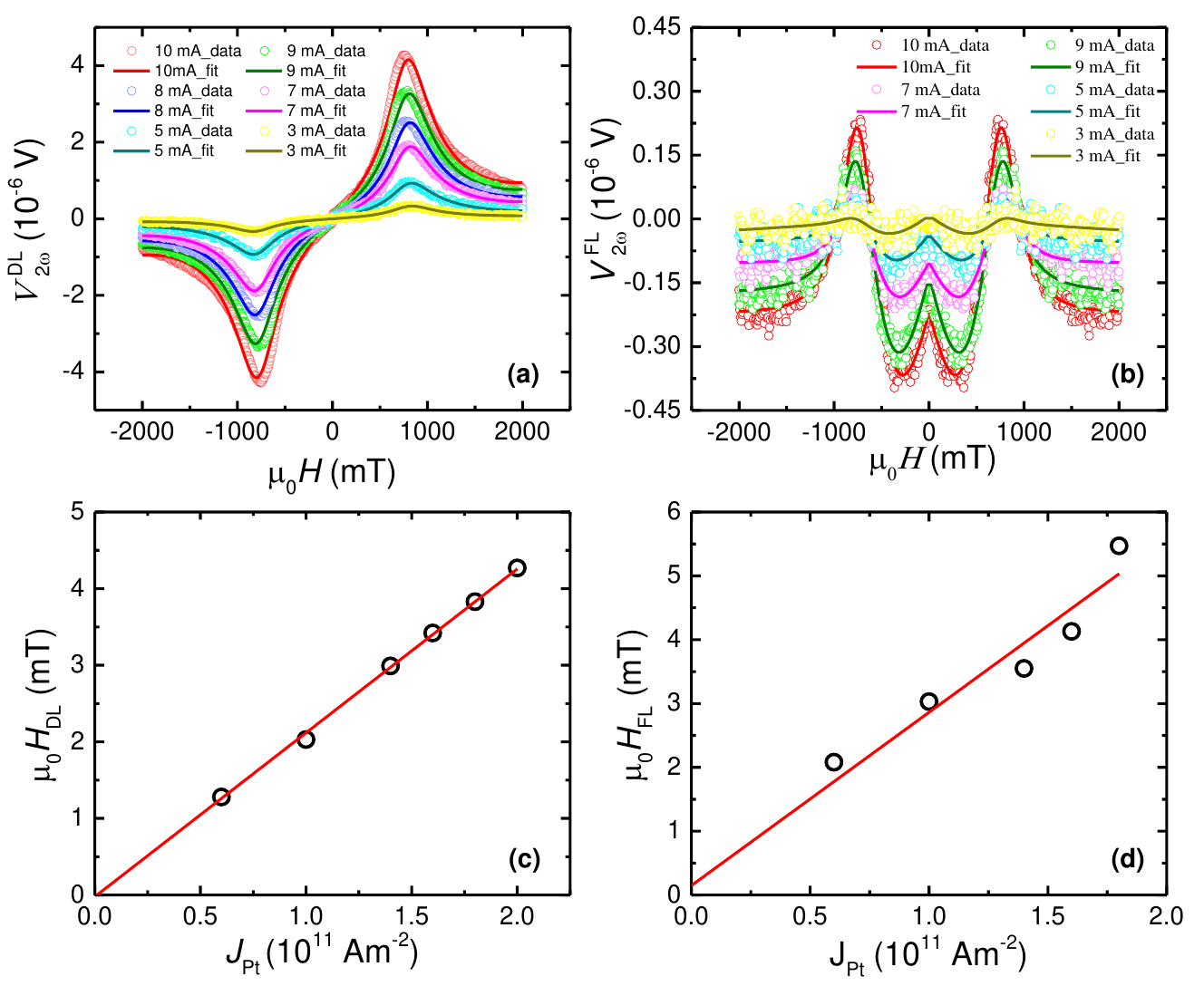}%
\caption{\label{fig:8}(a), (b) Second harmonic Hall voltages as a function of external magnetic field in the DL and FL geometry, respectively, for various input currents represented by different colors. The solid lines are fits to extract the DL torques. (c), (d) DL and FL effective fields vs $J_{\rm Pt}$ for the SL sample with $t_{\rm Tb}=0.25$\,nm.}
\end{figure}

Harmonic Hall voltage measurement technique has been used to quantify the damping-like (DL) torque in the SL samples. For the DL geometry, the current has been applied in-plane parallel to the external field. The Hall voltage can be represented as \cite{SOT_Hayashi,SOT_Wu}:
\begin{align}
    V_{\rm H} &= V_\omega\sin(\omega t)+V_{2\omega}\cos(2\omega t) \\
    V_\omega &= (R_0+R_{\rm A}\cos\theta+R_{\rm P}\sin^2\theta\sin2\phi)I_0 \\
    V_{2\omega} & \begin{multlined}[t]
        =\frac{1}{2}(R_{\rm A}\sin\theta-R_{\rm P}\sin2\theta\sin2\phi)\Delta\theta I_0 \\
        -R_{\rm P}\sin^2\theta\cos2\phi\Delta\phi I_0
    \end{multlined}
\end{align}
where $\theta$ and $\phi$ are the polar and azimuth angles of magnetization, respectively, $R_{\rm A}$ is the anomalous Hall resistance, $R_{\rm P}$ is the planar Hall resistance, and $R_0$ is a spurious resistance due to imperfection in the device. The ordinary Hall resistance is negligible. When an ac current of angular frequency $\omega$ is injected into the Hall bars, it generates effective SOT fields that produce magnetic oscillations around equilibrium. These oscillations modulate the Hall voltage in synchronization with the current. Notably, $V_\omega$ provides information about the magnetization direction in the presence of an external magnetic field, whereas $V_{2\omega}$ contains information about the magnetization oscillations generated by the torques. To extract the SOT effective field ($\mu_0H_{\rm DL}$), we simultaneously measure $V_\omega$ and $V_{2\omega}$ as a function of the in-plane magnetic field. The external magnetic field is applied slightly off-plane at an angle $\delta\theta\approx 0.12$\,rad ($\sim 7$ degrees) to ensure uniform magnetization inside the Hall bar. In DL geometry ($\theta = \frac{\pi}{2} + \delta\theta$, $\phi = 0$):
\begin{equation}
    \begin{multlined}[t]
        V_{2\omega}^{\rm DL}=-\frac{1}{2}\sin\theta\left[\frac{H_{\rm DL}R_{\rm A}}{H_{\rm K}\cos2\theta-H\sin(\delta\theta-\theta)}\right.\\
        \left.+\frac{2H_{\rm FL}R_{\rm P}\sec(\delta\theta)\sin\theta}{H}\right]I_0
    \end{multlined}
    \label{DL_fit}
\end{equation}
and, in field-like (FL) geometry ($\theta=\frac{\pi}{2}+\delta\theta$, $\phi=\frac{\pi}{2}$):
\begin{equation}
    \begin{multlined}[t]
        V_{2\omega}^{\rm FL}=\frac{1}{2}\sin\theta\cos\theta\left[\frac{H_{\rm FL}R_{\rm A}}{H_{\rm K}\cos2\theta-H\sin(\delta\theta-\theta)}\right.\\
        \left.+\frac{2H_{\rm DL}R_{\rm P}\sec(\delta\theta)\sin\theta}{H}\right]I_0
    \end{multlined}
    \label{FL_fit}
\end{equation}
$H_{\rm DL}$ and $H_{\rm FL}$ are the DL and FL effective fields, respectively. $R_{\rm A}$, $H_{\rm K}$, and $\theta$ are extracted from $V_\omega$. By fitting eqs.~\ref{DL_fit} and ~\ref{FL_fit} simultaneously as a function of $H_x$ and $H_y$, the $H_{\rm DL}$ and $H_{\rm FL}$ are extracted as a function of the $J_{\rm Pt}$.  Additionally, we note that $R_{\rm P}$ is one order of magnitude smaller than $R_{\rm A}$ and hence that the fits are dominated by the first terms of eqs.~\ref{DL_fit} and ~\ref{FL_fit}. The fitting of $V_{2\omega}$ in DL and FL geometry is shown in Fig.~\ref{fig:8}(a) and (b), respectively for the SL sample with $t_{\rm Tb}=$ 0.25 nm. Additionally, the variation of $H_{\rm DL}$ and $H_{\rm FL}$ as a function of $J_{\rm Pt}$ is shown in Fig.~\ref{fig:8}(c), and (d), respectively. From the best fit of Fig.~\ref{fig:8}(b), we obtain $\mu_0H_{\rm DL}=(2.14\pm0.15)\times10^{-11}$ mT\,A$^{-1}$m$^2$ and $\mu_0H_{\rm FL}=(2.84\pm0.19)\times10^{-11}$ mT\,A$^{-1}$m$^2$. Subsequently, the spin Hall angle $\theta_{\rm H}$ has been estimated by:
\begin{equation}
    \theta_{\rm H}=\frac{\mu_0H_{\rm DL}M_{\rm s}t_{\rm mag}}{\hbar/(2e) J_{\rm Pt}}\approx 0.086 \pm 0.006
    \label{SHA_fit}
\end{equation}
We note that while $\theta_{\rm H}$ can be extracted independently from equations ~\ref{conversion} and ~\ref{SHA_fit}, their forms are not identical due to the field representing two different physical quantities. The field $H$ in Eq.~\ref{conversion} represents the precessional term in the LLG equation whereas the DL effective field $H_{\rm DL}$ corresponds to the damping term. We note that the FL effective field is of the same order of magnitude as the DL field, questioning its being neglected in the analysis. The reason is that this field does not drive the domain walls, but competes with the DMI and the Néel wall demagnetizing fields. As for a typical current density at depinning of $5\times10^{11}$ A/m$^2$\, one has $\mu_0 H_{FL}=14$ mT\, it is clear that neglecting it is justified in first approximation.

\section{Joule heating and current induced DW dynamics}

A rough estimation of the temperature rise produced by Joule heating can be obtained by considering the heat diffusion through the $\rm SiO_{2}$ layer below the track \cite{TrackHeating, JouleHeating}. Indeed, the heat diffusion length is given by $l_D=\sqrt{Dt}$, where $D=\frac{K}{C\rho}$ is the heat diffusion coefficient, with $K$ the thermal conductivity, $C$ the specific hear and $\rho$ the density. For the values of $\rm SiO_{2}$ ($K=1.3 \rm W/m.K$, $C=680 \rm J/kg.K$, and $\rho=2.3 \rm g/cm^3$) and a typical current pulse duration $t=10 \rm ns$, the diffusion length $l_D=92 \rm nm$ is smaller than $\rm SiO_{2}$ layer thickness of 280 nm. Therefore, during the current pulse, the heat should propagate only just below the track without reaching significantly the Si substrate. The heated volume should be of the order of $V(t) \sim l_Dwl$ leading to a temperature rise $\Delta T \sim Pt/CV(t)$, where $P=R(Jwh)^2$ is the power dissipated ($w, h, l$ are the track width, thickness, and length, respectively). For the track resistance $R=240 \rm \Omega$ and typical current density $J=3.5\times10^{11}$ A$\rm m^{-2}$, the estimated temperature rise is $\Delta T(10 \rm ns) \sim 220K$. This large value essentially results from the very low heat conductivity of the $\rm SiO_{2}$ layer (which is two orders of magnitude lower than for Si and GaAs).

\bibliography{manuscript}

\end{document}